\documentstyle[12pt]{article}

\def\map{\mbox{\boldmath $\phi$}}
\def\be{\begin{equation}}
\def\ee{\end{equation}}
\def\la{\langle}
\def\ra{\rangle}
\def\IP{\hbox{\rm I\kern -1.6pt{\rm P}}}
\def\IC{{\hbox{\rm C\kern-.58em{\raise.53ex\hbox{$\scriptscriptstyle|$}}
    \kern-.55em{\raise.53ex\hbox{$\scriptscriptstyle|$}} }}}
\def\IN{\hbox{I\kern-.2em\hbox{N}}}
\def\IR{\hbox{\rm I\kern-.2em\hbox{\rm R}}}
\def\ZZ{\hbox{{\rm Z}\kern-.3em{\rm Z}}}
\def\IT{\hbox{\rm T\kern-.38em{\raise.415ex\hbox{$\scriptstyle|$}} }}
\def\notsub{\hbox{$\subset$\kern-.55em\hbox{/}}}

\newtheorem{theorem}{Theorem}
\newtheorem{lemma}[theorem]{Lemma}

\newtheorem{corollary}[theorem]{Corollary}

\begin{document}
\title{The existence of Burnett coefficients in the periodic Lorentz gas}
\author{
N. I. Chernov\thanks{Department of Mathematics, University of Alabama at
Birmingham, Birmingham, AL 35294.}\hspace{3mm}and
C. P. Dettmann\thanks{Center for Studies in Physics and Biology,
Rockefeller University, New York, NY 10021.}}



\maketitle

\begin{abstract} The linear super-Burnett coefficient gives
corrections to the diffusion equation in the form of higher
derivatives of the density.  Like the diffusion coefficient, it
can be expressed in terms of integrals of correlation functions,
but involving four different times.  The power law decay of
correlations in real gases (with many moving particles) and the
random Lorentz gas (with one moving particle and fixed scatterers)
are expected to cause the super-Burnett coefficient to diverge in
most cases. Here we show that the expression for the super-Burnett
coefficient of the periodic Lorentz gas converges as a result of
exponential decay of correlations and a nontrivial cancellation of
divergent contributions.
\end{abstract}

\section{Introduction}
To leading order, the equations of hydrodynamics describe the
transport of microscopically conserved quantities supplemented by
linear constitutive relations between the relevant thermodynamic forces
and fluxes.  For example, the conservation of the number of particles of
a light species moving in an arrangement of fixed scatterers is described
macroscopically using the continuity equation
\begin{equation}
\frac{\partial n}{\partial t}=-\nabla\cdot{\bf J}
\end{equation}
relating the number density $n$ to the current $\bf J$.  Under a hydrodynamic
approximation where $n$ is ``smooth'', hence all of its derivatives
are small, the flux ${\bf J}$ is given as a linear function of the force
$\nabla n$, that is, in an isotropic homogeneous medium,
\begin{equation}
{\bf J}=-D\nabla n
\end{equation}
leading to the diffusion equation
\begin{equation}
\frac{\partial n}{\partial t}=D\nabla^2n
\end{equation}
where $D$ is the diffusion coefficient.  Similar considerations involving
energy and momentum conservation lead to the Navier-Stokes equations for
a fluid.

There have been many attempts to go beyond this leading hydrodynamic
approximation.  The constitutive relation can contain an expansion in
terms of higher derivatives of the forces and/or higher powers of the forces.
In the former case one talks of linear Burnett~\cite{B} coefficients, and in
the latter, of nonlinear Burnett coefficients.  The first correction in either
case (coefficient of $\nabla^2n$ or $(\nabla n)^2$) is often called a
Burnett coefficient and the second correction (coefficient of $\nabla^3n$
or $(\nabla n)^3$) a super-Burnett coefficient, and so on.  We consider the
Lorentz gas, in which the light particles do not mutually interact, so
only linear relations are possible.  In addition, it is equally likely for
a particle to move in one direction or the opposite direction, so the
$\nabla^2n$ term cancels and we are left with the linear super-Burnett
coefficient.  We will dispense with this cumbersome terminology and call
it simply the Burnett coefficient.

Using the Boltzmann equation, hence a limit in which all collisions are
uncorrelated, the Chapman-Enskog method allows the computation of $D$ as
well as other and higher order transport coefficients of a low density gas.
One of the greatest surprises of nonequilibrium kinetic theory was the
discovery~\cite{DC} that density corrections due to recollisions of several
particles diverge.  This leads to a nonanalytic dependence of transport
coefficients on density in the case of the random Lorentz gas~\cite{LW},
where noninteracting point particles collide with a random arrangement of
non-overlapping hard spheres in two or three dimensions,
but not in an exactly solvable one dimensional model~\cite{LP}.

Subsequently the divergences were related to a power law
decay of correlations (see~\cite{EW} and references therein), the
``long time tails'', which are
due to density fluctuations in infinite random systems.  Since transport
coefficients can be computed using integrals of correlation functions
(see below), a slow decay of correlations can lead to divergences.
As discussed in a review of random Lorentz models~\cite{vB}, the diffusion
coefficient is typically defined, but Burnett coefficients may not be
since the latter have more divergent expressions in general.  The degree of
divergence is greater for real gases than for Lorentz gases, and decreases
with the dimension of the system.  Here we show that, as a result of the
exponential decay of correlations in the periodic Lorentz model, the
combination of four-time correlation functions used to compute the Burnett
coefficient decays sufficiently rapidly to ensure convergence.  Thus a Burnett
coefficient is defined for a random Lorentz gas in a dilute approximation
and for a periodic Lorentz gas under general conditions.

\section{The Burnett coefficient}
Our notation follows that of Gaspard~\cite{G} who gives an account of
deterministic diffusion including expressions for the Burnett coefficients.
He suggests stretched exponential bounds on the four time correlations
which we find, but does not give the precise form or rigorous proof.

We consider a periodic Lorentz gas in which a point particle
moving with unit velocity undergoes specular reflections with a
periodic array of scatterers.  We consider only the case of finite
horizon, so that time between collisions is bounded.  The most
usual case is that of circular scatterers in a hexagonal lattice,
but we demand only that the scatterers be convex and sufficiently
smooth, specifically $C^3$.  We allow for any dimension $d\geq 2$,
but note that a single spherical scatterer per unit cell cannot
satisfy the finite horizon condition for $d>2$.  Thus we need
either nonspherical scatterers or more than one per unit cell when
$d>2$.

Because the scatterers lie on a periodic lattice, it is possible to represent
the dynamics in terms of a map $\map$ acting at the surface of the scatterers.
We denote the position and velocity direction at the surface of a scatterer
in the elementary cell by ${\bf x}$.  Then the lattice translation vector
${\bf a}({\bf x})$ gives the lattice displacement associated with the free
flight following a collision at ${\bf x}$; this is a linear combination of the
lattice basis vectors ${\bf e}^{(\alpha)}$ ($\alpha=1\ldots d$) with integer
coefficients and is a piecewise constant function of ${\bf x}$. $T({\bf x})$
is the time for this free flight.  We can
compute time averages, which for almost all initial conditions are equivalent
to the natural measure on the surface of the scatterers $\langle\rangle$
due to ergodicity.  We define $\Delta T({\bf x})=T({\bf x})-\langle T\rangle$
so that $\langle\Delta T\rangle=0$.  We use Greek indices to refer to spatial
components such as $a_\alpha$, and Latin indices to refer to collisions,
with the shorthand notation $a_\alpha^i=a_\alpha(\map^i{\bf x})$ and
$\Delta T^i=\Delta T(\map^i{\bf x})$.  We have $\langle{a_\alpha^i}\rangle=0$
and also odd combinations such as
$\sum_{ij=-\infty}^{\infty}\langle a_\alpha^0 a_\beta^i a_\gamma^j\rangle=0$
since time reversibility
and phase space volume conservation imply that there are equal and opposite
contributions from time reversed trajectory segments.

The general periodic Lorentz gas is homogeneous but not isotropic at large
scales, so the hydrodynamic equation contains transport coefficients that
are fully symmetric tensors, but do not depend on position:
\begin{equation}
\frac{\partial n}{\partial t}=D_{\alpha\beta}\nabla_\alpha\nabla_\beta n
+B_{\alpha\beta\gamma\delta}
\nabla_\alpha\nabla_\beta\nabla_\gamma\nabla_\delta n+\ldots\;\;.
\end{equation}
where $B_{\alpha\beta\gamma\delta}$ are the Burnett coefficients and repeated
Greek indices are summed.
A general solution can be obtained with a Fourier-Laplace transform of the
form $n\sim\exp(st+i{\bf k}\cdot{\bf r})$ leading to a dispersion relation
\begin{equation}\label{e:disp}
s({\bf k})=-D_{\alpha\beta}k_\alpha k_\beta
+B_{\alpha\beta\gamma\delta}k_\alpha k_\beta k_\gamma k_\delta+\ldots
\end{equation}
The Green-Kubo relations for the transport coefficients in terms
of integrals (sums) of correlation functions are given in
continuous time in, for example, Ref.~\cite{vB}, and in discrete
time~\cite{G} they are
\begin{equation}
D_{\alpha\beta}=\frac{1}{2\langle T\rangle}
\sum_{i=-\infty}^{\infty}\langle a_\alpha^0 a_\beta^i\rangle
\end{equation}
and
\begin{eqnarray}
B_{\alpha\beta\gamma\delta}&=&\frac{1}{24}\left[F_{\alpha\beta\gamma\delta}
+4C(D_{\alpha\beta}D_{\gamma\delta}+D_{\alpha\gamma}D_{\beta\delta}
+D_{\alpha\delta}D_{\beta\gamma})\right.\\\nonumber
&&\left.\!\!\!\!\!\!\!\!
-2(D_{\alpha\beta}E_{\gamma\delta}+D_{\alpha\gamma}E_{\beta\delta}
+D_{\alpha\delta}E_{\beta\gamma}+D_{\beta\gamma}E_{\alpha\delta}
+D_{\beta\delta}E_{\alpha\gamma}+D_{\gamma\delta}E_{\alpha\beta})\right]
\end{eqnarray}
with
\begin{eqnarray}
C&=&\frac{1}{\langle T\rangle}\sum_{i=-\infty}^{\infty}
\langle\Delta T^0\Delta T^i\rangle\\
E_{\alpha\beta}&=&\frac{1}{\langle T\rangle}\sum_{i,j=-\infty}^{\infty}
\langle\Delta T^0 a_\alpha^i a_\beta^j\rangle\\
F_{\alpha\beta\gamma\delta}&=&\frac{1}{\langle T\rangle}
\sum_{i,j,k=-\infty}^{\infty}\left[
\langle a_\alpha^0 a_\beta^i a_\gamma^j a_\delta^k \rangle
-\langle a_\alpha^0 a_\beta^i \rangle \langle a_\gamma^j a_\delta^k \rangle
\right.\nonumber\\&&\left.
-\langle a_\alpha^0 a_\gamma^j \rangle \langle a_\beta^i a_\delta^k \rangle
-\langle a_\alpha^0 a_\delta^k \rangle \langle a_\beta^i a_\gamma^j \rangle
\right]
\end{eqnarray}
It is possible to make the lower limits of the sums all zero with some extra
symmetrization and special treatment of the zero terms.  The convergence
of the sum for $C$ follows in a straightforward manner from the known
exponential decay of correlations in this system~\cite{Ch99}.  The convergence
of the sum for $E_{\alpha\beta}$ follows from Theorem 3 in the following
section.  The sum for $F_{\alpha\beta\gamma\delta}$ is still less clear since
each term diverges; however Theorem 1 in the following section
show that the combination converges, leading to finite Burnett coefficients.

In the case of the hexagonal Lorentz gas, the six-fold rotational symmetry
is sufficient to restrict the coefficients to their isotropic forms
$D_{\alpha\beta}=D\delta_{\alpha\beta}$ and
$B_{\alpha\beta\gamma\delta}=B(\delta_{\alpha\beta}\delta_{\gamma\delta}
+\delta_{\alpha\gamma}\delta_{\beta\delta}
+\delta_{\alpha\delta}\delta_{\beta\gamma})/3$.  It seems likely that a
straightforward application of our results could also show the existence
of the higher order Burnett coefficients, for which there exist
long expressions similar to those given here.  In this case it would be
interesting to consider the radius of convergence of Eq.~(\ref{e:disp}).

\section{Proofs}

The billiard map $\map$ is defined on the collision space $M$ of
the periodic Lorentz gas. In the case of finite horizon, the map
$\map$ and the above functions $a_{\alpha}$ and $\Delta T$ are
piecewise H\"older continuous \cite{BSC91,Ch94}. This means that
$M$ can be partitioned into finitely many domains with piecewise
smooth boundaries on each of which the function is H\"older
continuous.

We always denote by $\la \cdot\ra$ the integration over $M$ with
respect to the invariant equilibrium measure. In all theorems
below, $f_0,f_1,\ldots$ denote piecewise H\"older continuous
functions on $M$ such that $\la f_k\ra=0$ for all $k$. For $i\geq
0$ we put $f_k^i=f_k\circ \map^i$.

\begin{theorem}
The following series
\be
    \sum_{i_1,i_2,i_3=0}^{\infty}\la f_0^0f_1^{i_1}f_2^{i_2}f_3^{i_3}\ra
    -\la f_0^0f_1^{i_1}\ra\la f_2^{i_2}f_3^{i_3}\ra
    -\la f_0^0f_2^{i_2}\ra\la f_1^{i_1}f_3^{i_3}\ra
    -\la f_0^0f_3^{i_3}\ra\la f_1^{i_1}f_2^{i_2}\ra
       \label{cor4}
\ee
always converges absolutely. \label{tm0}
\end{theorem}

We prove general estimates on multiple correlation functions, from
which the above theorem will follow.

\begin{theorem}
Let $i_1\leq\cdots \leq i_k$ and $1\leq t\leq k-1$. Then
\be
    |\la f_1^{i_1}\cdots f_k^{i_k}\ra -
    \la f_1^{i_1}\cdots f_t^{i_t}\ra\la f_{t+1}^{i_{t+1}}\cdots f_k^{i_k}\ra|
    \leq C_k\cdot |i_k-i_1|^2\,\lambda^{|i_{t+1}-i_t|^{1/2}}
       \label{F3}
\ee
where $C_k>0$ depends on the functions $f_1,\ldots,f_k$, and
$\lambda<1$ is independent of $k$ and $f_1,\ldots,f_k$.
\label{tm1}
\end{theorem}

This allows us to ``decouple'' multiple correlations provided the
gap between the time moments $i_t$ and $i_{t+1}$ is long enough.
\medskip

\noindent {\em Remark}. In view of recent advances in the study of
correlations for billiard dynamics \cite{LSY98,Ch99} it is very
likely that the above estimate can be upgraded to
$C_k\lambda^{|i_{t+1}-i_t|}$ for some $\lambda<1$, which would be
an exponential decay of (multiple) correlations. However, the
estimate (\ref{F3}) is easier to prove, and for our purposes it is
good enough.

We postpone the proof of the Theorem~\ref{tm1} and first obtain
Theorem~\ref{tm0} and other interesting implications.

\begin{theorem}
Let $i_1\leq\cdots \leq i_k$. Consider the $k-1$ differences
$m_t=i_{t+1}-i_t$ for $1\leq t\leq k-1$. Let $m_{(1)}\leq
m_{(2)}\leq\cdots\leq m_{(k-1)}$ be the ordered differences
$m_1,\ldots,m_{k-1}$. Let $m=m_{(r)}$ where $r=[(k+1)/2]$. Then
$$
    |\la f_1^{i_1}\cdots f_k^{i_k}\ra|\leq C_k\cdot |i_k-i_1|^2\,\lambda^{m^{1/2}}
$$
where $C_k>0$ and $\lambda<1$ (here $\lambda$ is the same as in
the previous theorem). \label{tm2}
\end{theorem}

To prove this theorem, we use a simple lemma:

\begin{lemma}
Let $k\geq 2$ and $r=[(k+1)/2]$. If we partition $k$ objects into
$k-r+1$ nonempty groups, then at least one group contains exactly
one object. \label{lmobj}
\end{lemma}

Now, Theorem~\ref{tm2} is proved by applying Theorem~\ref{tm1} to
the $k-r$ largest gaps $m_{(k-1)},\ldots, m_{(r)}$ and recalling
that $\la f_j\ra=0$ for every $j$.

Note that there is no similar bound on multiple correlations in
terms of the next difference, $m_{(r+1)}$. Indeed, no matter how
large $m_{(r+1)}$ is, we can always arrange the time moments
$i_1,\ldots,i_k$ so that $i_1=i_2$, $i_3=i_4$, $\ldots$,
$i_{k-1}=i_k$, and then the correlation $\la f_1^{i_1}\cdots
f_k^{i_k}\ra$ need not be even small, in general it will stay
bounded away from zero.

\begin{corollary}
For $k=2$, we obtain the known \cite{BSC91} bound on the
2-correlation function:
$$
    |\la f_1^{i_1}f_2^{i_2}\ra|\leq C(f_1,f_2)\cdot\lambda_0^{|i_2-i_1|^{1/2}}
$$
for some $C(f_1,f_2)>0$ and $\lambda_0<1$.
\end{corollary}

We now prove Theorem~\ref{tm0}. We can assume that $i_1\leq
i_2\leq i_3$ in (\ref{cor4}) and put
$\bar{m}=\max\{i_1,i_2-i_1,i_3-i_2\}$. Then it easily follows from
Theorem~\ref{tm1} and the above corollary that $$
   |\la f_0^0f_1^{i_1}f_2^{i_2}f_3^{i_3}\ra
    -\la f_0^0f_1^{i_1}\ra\la f_2^{i_2}f_3^{i_3}\ra
    -\la f_0^0f_2^{i_2}\ra\la f_1^{i_1}f_3^{i_3}\ra
    -\la f_0^0f_3^{i_3}\ra\la f_1^{i_1}f_2^{i_2}\ra|
    \leq C\, \bar{m}^2\lambda^{\bar{m}^{1/2}}
$$
This easily implies the absolute convergence of the series
(\ref{cor4}). Theorem~\ref{tm0} is proved.

\begin{theorem}
Let $S_n=f_0^0+f_0^1+\cdots +f_0^{n-1}$. Then the $k$-th moment of
$S_n$ can be bounded by
$$
    |\la S_n^k \ra|\leq C(f_0,k)\cdot n^{[k/2]}
$$
where $C(f_0,k)>0$. \label{tm3}
\end{theorem}

This theorem follows from Theorem~\ref{tm2}. In the case $k=4$ see
the proof in \cite{BSC91}. The argument generalizes to any $k\geq
3$ rather directly. In view of the remark made after
Lemma~\ref{lmobj}, the above estimate cannot be improved, in
general. \medskip

{\em Proof of Theorem~\ref{tm1}}. We use the partition method
developed in \cite{BSC91} and further enhanced in \cite{Ch95}.
Note that \cite{BSC91,Ch95} discuss two-dimensional periodic
Lorentz gases, but all the results hold in any dimensions as well
\cite{Ch94}. We recall the essentials of the partition method in a
more general setting than our Lorentz gas model.

Let $\map:M\to M$ be a measurable transformation of a metric space
$M$ preserving a nonatomic Borel probability measure $\mu$. Let
${\cal A}=\{A_1,A_2,\ldots\}$ be a finite or countable measurable
partition of $M$, we denote diam$\, {\cal A} =\sup_i\{{\rm diam}\,
A_i\}$. We put ${\cal A}_n= \map^{-n}{\cal A}=\{\map^{-n}A_i\}$
for $n\geq 0$ and ${\cal A}_{n,k} ={\cal A}_n\vee\cdots\vee {\cal
A}_k$ for $k\geq n\geq 0$. A measure of dependence between two
partitions ${\cal A}= \{A_i\}$ and ${\cal B}=\{B_j\}$ of the space
$M$ is defined to be
$$
    \beta ({\cal A},{\cal B})=\sum_{i,j}|\mu(A_i\cap B_j)-
      \mu(A_i)\mu(B_j)|
$$
Based on this measure, we put
\be
     \beta_N(n)=\max_{0\leq k\leq N-n-1}
     \beta ({\cal A}_{0,k},{\cal A}_{k+n,N-1}),
         \label{betaNn}
\ee
for any $N\geq n\geq 0$. Note that $\beta_N(n)$ is called the
mixing coefficients of the partition $\cal A$, cf. \cite{Ch95}.

For any measurable bounded function $f$ on $M$ and a partition
$\cal A$ of $M$ we denote $\bar{f}_{\cal A}= \langle f|{\cal
A}\rangle$ the conditional expectation of $f$ with respect to the
$\sigma$-algebra generated by $\cal A$, and $\Delta_{\cal
A}f=f-\bar{f}_{\cal A}$. We put
$$
   {\cal H}_f(d)=\sup_{{\rm diam}\,{\cal A}\leq d}
    ||\Delta_{\cal A}f||_2
$$
Clearly, ${\cal H}_f(d)$ monotonically decreases to zero as
$d\to 0$, and the speed of decrease characterizes the
``smoothness'' of $f$, see \cite{Ch95} and below.

\begin{lemma}
For any functions $f_1,\ldots,f_k$ on $M$ and any partition $\cal
A$ with $d=\,{\rm diam}\,{\cal A}$ we have
\be
    |\la f_1^{i_1}\cdots f_k^{i_k}\ra -
    \la f_1^{i_1}\cdots f_t^{i_t}\ra\la f_{t+1}^{i_{t+1}}\cdots f_k^{i_k}\ra|\leq
    C_k\, \left[\sum_{t=1}^k{\cal H}_{f_t}(d)+\beta_{i_k-i_1}(i_{t+1}-i_t)\right]
      \label{cork}
\ee
\label{lmcor}
\end{lemma}

{\em Proof}. First, we replace $f_t$ by
$\bar{f}_t=\bar{f}_{t,{\cal A}}$. The error can be estimated by
$$
    |\la f_1^{i_1}\cdots f_k^{i_k}\ra -
    \la\bar{f}_1^{i_1}\cdots {\bar f}_k^{i_k}\ra|
     \leq C_k \sum_t||\Delta_{\cal A}f_t||_2
     \leq C_k \sum_t{\cal H}_{f_t}(d)
$$
which is a simple calculation using Schwarz' inequality, cf.
\cite{Ch95}. Two other correlations in (\ref{cork}) can be handled
similarly.

Next, it is rather straightforward that
$$
    |\la \bar{f}_1^{i_1}\cdots \bar{f}_k^{i_k}\ra -
    \la \bar{f}_1^{i_1}\cdots \bar{f}_t^{i_t}\ra
    \la \bar{f}_{t+1}^{i_{t+1}}\cdots \bar{f}_k^{i_k}\ra|\leq
    \prod_t||f_t||_{\infty}\times\beta_{i_k-i_1}(i_{t+1}-i_t)
$$
(for $k=2$, this was done in \cite{Ch95}, the general case is
similar). Lemma is proved. $\Box$\medskip

We now need a good partition $\cal A$ for our Lorentz gas map
$\map:M\to M$. Such a partition was constructed in
\cite{BSC91,Ch94} and further refined in \cite{Ch95}. Here we
state the result:

\begin{theorem}[see \cite{Ch95}]
Let $\map:M\to M$ be the billiard map for a periodic Lorentz gas
with finite horizon. Then for any $N\geq m\geq 1$ there is a
partition ${\cal A}={\cal A}_{N,m}$ such that \\ {\rm (i)} ${\rm
diam}\,{\cal A}\leq c_1\lambda_1^{m^{1/2}}$,\\ {\rm (ii)}
$\beta_N(m)\leq c_2N^2\lambda_2^{m^{1/2}}$\\ for some constants
$c_1,c_2>0$ and $\lambda_1,\lambda_2<1$. \label{tmpar}
\end{theorem}

It remains to estimate the term ${\cal H}_{f_t}(d)$ in
(\ref{cork}). Let $f$ be a piecewise H\"older continuous function
on $M$. Then one can easily obtain \cite{Ch95} that ${\cal
H}_f(d)\leq C(f)\cdot d^{\alpha}$ for some $\alpha>0$. Now
Theorem~\ref{tm1} follows from Lemma~\ref{lmcor} and
Theorem~\ref{tmpar}. \medskip

{\bf Acknowledgements}.  CD is grateful for discussions with E. G.
D. Cohen and J. R. Dorfman, and for the support of the Engineering
Research Program of the Office of Basic Energy Sciences at the US
Department of Energy, contract \#DE-FG02-88-ER13847. NC is
partially supported by NSF grant DMS-9732728.

\end{document}